\documentclass[twocolumn,smallextended,english,epjc3]{svjour3}
\usepackage[dvips]{graphicx}
\usepackage{amsmath}
\usepackage{graphicx}
\usepackage{amsfonts}
\usepackage{amssymb}
\usepackage{epsfig}
\usepackage{color}

\begin{document}
\title{ Greybody factor for black string in dRGT massive gravity}

\author{P. Boonserm \thanksref{ePB,add1,add2} \and  T. Ngampitipan \thanksref{eTN,add3}
\and Pitayuth Wongjun \thanksref{ePW,add2,add4}}

\thankstext{ePB}{e-mail: petarpa.boonserm@gmail.com}
\thankstext{eTN}{e-mail: tritos.ngampitipan@gmail.com}
\thankstext{ePW}{e-mail: pitbaa@gmail.com}

\institute{
 Department of Mathematics and Computer Science, Faculty of Science, Chulalongkorn University, Bangkok 10330, Thailand\label{add1}\,\,\and Thailand Center of Excellence in Physics, Ministry of Education, Bangkok 10400 Thailand\label{add2}\,\,\and Faculty of Science, Chandrakasem Rajabhat University,  Bangkok 10900, Thailand\label{add3}
\,\,\and The Institute for Fundamental Study, Naresuan University, Phitsanulok 65000, Thailand\label{add4}
}

\maketitle

\begin{abstract}
The greybody factor from the black string in the de Rham-Gabadadze-Tolley (dRGT) massive gravity theory is investigated in this study. The dRGT massive gravity theory is one of the modified gravity theories used in explaining the current acceleration in the expansion of the universe. Through the use of cylindrical symmetry, black strings in dRGT massive gravity are shown to exist. When quantum effects are taken into account, black strings can emit thermal radiation called Hawking radiation. The Hawking radiation at spatial infinity differs from that at the source by the so-called greybody factor. In this paper, we examined the rigorous bounds on the greybody factors from the dRGT black strings. The results show that the greybody factor crucially depends on the shape of the potential which is characterized by model parameters. The results agree with ones in quantum mechanics; the higher the potential, the harder it is for the waves to penetrate, and also the lower the bound for the rigorous bounds. 
\end{abstract}

%\noindent\textbf{keywords}: black string, cylindrical symmetry, dRGT massive gravity, greybody factor, rigorous bound

\section{Introduction}
Based on cosmological observations, our universe is expanding with an acceleration \cite{Supernova,Supernova2}. However, the explanation for this phenomenon remains unclear. Many authors propose the existence of exotic matter called dark energy to explain this observed cosmic acceleration. On the other hand, some authors modify gravity without dark energy. One of the modifications of gravity is to give mass to graviton. The successful and viable models of massive gravity are the de Rham–-Gabadadze–-Tolley (dRGT) models \cite{dRGT,dRGT2}. The reviews of the theory of massive gravity can be found in \cite{Hinterbichler,deRham:2014zqa}.
For spherical symmetry, the black hole solutions have also been found, and their thermodynamics properties extensively investigated \cite{Vegh:2013sk,Cai:2014znn,Ghosh:2015cva,Adams:2014vza,Xu:2015rfa,Nieuwenhuizen:2011sq,Brito:2013xaa,Berezhiani:2011mt,Cai:2012db,Babichev:2014fka,Volkov:2013roa,Babichev:2015xha,Capela:2011mh,Volkov:2012wp,Hu:2016hpm,Hu:2016mym,Zou:2016sab,Hendi:2017arn,Hendi:2017ibm,Hendi:2016usw,EslamPanah:2016pgc,Hendi:2016hbe,Hendi:2016uni,Hendi:2016yof,Arraut:2014uza,Arraut:2014iba,Arraut:2014sja,Kodama:2013rea}.

When quantum effects are taken into account, black holes can emit thermal radiation called Hawking radiation \cite{Hawking}. The original Hawking radiation emitted from a black hole is a blackbody radiation. Due to the curvature of spacetime, the Hawking radiation is modified, while propagating to spatial infinity. The radiation at spatial infinity differs from that at the emitter by the so-called greybody factor. There are various methods to find the greybody factors such as the matching technique and the WKB approximation \cite{Parikh,Fleming,Fernando,Lange,Kim,Escobedo,Harmark,Kanti,Dong}. Another interesting method is to bound the greybody factor from below \cite{1D,Bogo,Sch,phd,non,dirty,KN,MP,Tphd,china,epjc18}.

Besides the solution in the spherical symmetry, the solution to the Einstein field equation in the cylindrical symmetry has also been investigated and is known as the black string solution \cite{Lemos:1994xp,Lemos:1994fn}. This solution can be achieved by introducing the cosmological constant into the Einstein field equation. The charge and the rotating black string solutions can also be found \cite{Lemos:1995cm}. The quasinormal modes \cite{Cardoso:2001vs} and the greybody factor of the black string  have been investigated \cite{Ahmed}.

As we know, the dRGT massive gravity theory can provide a more general solution than the Schwarzschild-dS/AdS. Therefore, it is possible to obtain the cylindrical solution in the dRGT massive gravity theory \cite{Ghosh:2017cva}. The rotating solutions and their thermodynamic properties are also investigated \cite{prepare}. The quasinormal mode for the dRGT black string solution have been investigated as well \cite{Ponglertsakul:2018smo}, while the greybody factor have not been investigated yet. In the present work, the rigorous bounds on the greybody factor from the dRGT black strings are examined.

This paper is organized as follows. In Sec. \ref{back}, the background of the dRGT black string is provided. The horizon structures are analyzed in Sec. \ref{horizon}. The equation of motion of the massless scalar field emitted from a dRGT black hole and the gravitational potential which modifies the scalar field are derived in Sec. \ref{eom}. The rigorous bounds on the greybody factors are calculated in Sec. \ref{bound}, and the conclusions are given in Sec. \ref{con}.

\section{dRGT black string background}\label{back}
In this section, the dRGT massive gravity theory, including how  the black string solution can be obtained, is roughly reviewed. The main concept in the modification of the general relativity in the dRGT massive gravity is the addition of a suitable graviton mass to General Relativity (GR), of which the action can be written as
\cite{dRGT,dRGT2}
\begin{eqnarray}\label{action}
 S = \int d^4x \sqrt{-g}\; \frac{1}{2} \left[ R(g) +m_g^2\,\, {\cal U}(g, f)\right],
\end{eqnarray}
where $R$ is the Ricci scalar, ${\cal U}$ is a potential term used in characterizing the behaviour of the mass term of graviton, and $m_g$ is the parameter interpreted as the graviton mass. The suitable form of the potential ${\cal U}$ in four-dimensional spacetime is given by
\begin{eqnarray}\label{potential}
 {\cal U}(g, \phi^a) &=& {\cal U}_2 + \alpha_3{\cal U}_3 +\alpha_4{\cal U}_4 ,\\
  {\cal U}_2&\equiv&[{\cal K}]^2-[{\cal K}^2] ,\\
 {\cal U}_3&\equiv&[{\cal K}]^3-3[{\cal K}][{\cal K}^2]+2[{\cal K}^3] ,\\
 {\cal U}_4&\equiv&[{\cal K}]^4-6[{\cal K}]^2[{\cal K}^2]+8[{\cal K}][{\cal
K}^3]\nonumber \\
&& +3[{\cal K}^2]^2-6[{\cal K}^4],
\end{eqnarray}
where $\alpha_3$ and $\alpha_4$ are dimensionless free parameters of the theory. The quantity $[{\cal K}]$ denotes the trace of the metric ${\cal K}^\mu_\nu$, defined by
\begin{eqnarray}
 {\cal K}^\mu_\nu =
\delta^\mu_\nu-\sqrt{g^{\mu\rho}f_{\rho\nu}}, \label{K-tensor}
\end{eqnarray}
where  $[{\cal K}^n]=({\cal K}^n)^\mu_\mu$ and $({\cal K}^n)^\mu_\nu = {\cal K}^\mu_{\rho_2} {\cal K}^{\rho_2}_{\rho_3} ...{\cal K}^{\rho_{(n-1)}}_{\rho_n} {\cal K}^{\rho_n}_{\nu} $ for $n \geq 2$. It is important to note that the potential terms include the non-dynamical metric $f_{\mu\nu}$ called the fiducial metric or the reference metric. The form of solution of the physical metric $g_{\mu\nu}$ significantly depends on the form of the fiducial metric \cite{Chullaphan:2015ija,Tannukij:2015wmn,Nakarachinda:2017oyc}. The corresponding equation of motion to the above action can be written as
\begin{eqnarray}\label{EoM}
 G_{\mu\nu} +m_g^2 X_{\mu\nu} = 0, \label{modEFE}
\end{eqnarray}
where
\begin{eqnarray}
 X_{\mu\nu} &=& {\cal K}_ {\mu\nu} -{\cal K}g_ {\mu\nu} -\alpha\left({\cal K}^2_{\mu\nu}-{\cal K}{\cal K}_{\mu\nu} +\frac{{\cal U}_2}{2}g_{\mu\nu}\right) \nonumber \\
 &&+3\beta\left( {\cal K}^3_{\mu\nu} -{\cal K}{\cal K}^2_{\mu\nu} +\frac{{\cal U}_2}{2}{\cal K}_{\mu\nu} - \frac{{\cal U}_3}{6}g_{\mu\nu} \right), \,\,\,\,\,\,\label{effemt}\\
  \alpha_3 &=& \frac{\alpha-1}{3}~,~~\alpha_4 =
\frac{\beta}{4}+\frac{1-\alpha}{12}.
\end{eqnarray}
Due to the existence of the Bianchi identity, the tensor $X_{\mu\nu}$ obeys the covariant conservation as
\begin{eqnarray}\label{BiEoM}
 \nabla^\mu X_{\mu\nu} = 0.
\end{eqnarray}
By imposing the static and cylindrical symmetry, a general form of the black string solution (physical metric) can be written as \cite{Ghosh:2017cva}
\begin{eqnarray}
ds^{2} = -f(r)dt^{2} + \frac{dr^{2}}{f(r)} + r^{2}(d\varphi^2 + \alpha_{g}^2dz^2), \label{solutiong}
\end{eqnarray}
where $\alpha_g$ is a constant. By choosing the form of the fiducial metric as
\begin{eqnarray}\label{fiducial metric}
f_{\mu\nu}=\text{diag}(0,0,h_0^2  , h_0^2 ),
\end{eqnarray}
where $h_0$ is a constant, the function $f(r)$ in the physical metric can be written as \cite{Ghosh:2017cva}
\begin{equation}
f(r) = c_2 r^{2} - \frac{4M}{r} - c_{1}r + c_{0} \label{f}
\end{equation}
and $d\Omega^{2} = d\varphi^2 + \alpha_{g}^2dz^2$, $M=\bar{M}/\alpha_{g}$, where $\bar{M}$ is the ADM mass per unit length in the z direction. The parameters above can be written in terms of the original parameters as
\begin{subequations}
\begin{eqnarray}
c_2 &=& m^2_g \left(1+\alpha+\beta\right),\\
c_{1} &=& m^2_g h_{0}(1 + 2\alpha + 3\beta),\\
c_{0} &=& m^2_g h_{0}^{2}(\alpha + 3\beta).
\end{eqnarray} \label{pardef}
\end{subequations}

The solution in Eq. (\ref{solutiong}), including function $f$ in Eq. (\ref{f}), is an exact black string solution in dRGT massive gravity which, within the limit $ c_2 = \alpha_g^2$ and $c_0=c_1=0$, naturally goes over to Lemos's black string in GR with cosmological constant \cite{Lemos:1994xp,Lemos:1994fn} . In particular, it incorporates the cosmological constant term ($c_2$ term) naturally in terms of the graviton mass. Moreover, this solution also provides a global monopole term ($c_0$ term) as well as another non-linear scale term ($c_1$ term).

It is important to note that the strong coupling scale of the dRGT massive gravity theory is $\Lambda_3^{-1} = (m_g^2 M_{Pl})^{1/3} \sim 10^{3} ~\text{km} \ll  r_V \sim 10^{16} ~\text{km}$ so that we do not have to worry about the strong coupling issue in dRGT massive gravity for a system of scale below $\Lambda_3$ (or of length scale beyond $\sim10^3$ km), where $r_V$ is the Vainshtein radius characterized by the non-linear scale of the massive gravity theory \cite{Ghosh:2017cva}.

One can see that the horizon structure depends on the sign of $c_2$. If $c_2>0$, corresponding to the Anti de Sitter-like solution, the maximum number of horizons are three. If $c_2<0$, corresponding to the de Sitter-like solution, the maximum number of horizons are two. This behaviour is explicitly shown in the next section.

\section{Horizon structure}\label{horizon}
In order to investigate the structure of the horizons for the solution in Eq. (\ref{solutiong}), where $f$ is in Eq. (\ref{f}), one has to find the number of possible extremum points. As a result, this depends on the asymptotic behaviour of the solution. For the asymptotic dS solution, $c_2 < 0$, the solution becomes the dS black string for the large-$r$ limit, while the solution becomes AdS black string for the large-$r$ limit of the asymptotic AdS solution, $c_2 > 0$. As a result, one can find the conditions to obtain one positive real maximum of $f$ for the asymptotic dS solution. For the asymptotic AdS solution, one can find the conditions to have one positive real maximum and one negative real minimum $f$. We will investigate this behaviour separately in the following subsection.

It is important to note that by choosing the fiducial metric as $h_0 = 0$, the solution becomes AdS/dS black string solution. This is not surprising since the potential term becomes a constant.

\subsection{Asymptotic dS solution}
For the asymptotic dS solution, $c_2 < 0$, one can find conditions for having two horizons by solving $f'=0$ to obtain a real positive value of the radius, $r$ by
\begin{equation}
r_{ex}= (\sqrt{3}-1)\left(\frac{M}{-c_2}\right)^{1/3}.
\end{equation}
Note that to guarantee this existence, we choose the condition on $c_1$ as $c_1 = 6 (M c_2^2)^{1/3}$. As a result, $f$ at the extremum can be written as
\begin{equation}
f(r_{ex}) = c_0 +6 \sqrt{3} c_2 \left(-\frac{M}{c_2}\right)^{2/3}.
\end{equation}
In order to have two horizons, $f(r_{ex})$ must be positive. Therefore, let us define the parameter $\beta_m > 1$ for having two horizons as
\begin{equation}
 c_0 = -\beta_m  6 \sqrt{3} c_2 \left(-\frac{M}{c_2}\right)^{2/3}.
\end{equation}
By substituting this parameter into $f$, and then finding the solution of $f =0$ for $r$, one obtains two horizons as follows
\begin{eqnarray}
r_1 &=& \frac{2 (c_2^2 M)^{1/3}}{c_2}\left( 1+ \sqrt{X} \cos\left(\frac{\cos^{-1}Y}{3} + \frac{\pi}{3}\right) \right),\\
r_2 &=& \frac{2 (c_2^2 M)^{1/3}}{c_2}\left( 1- \sqrt{X} \cos\left(\frac{\cos^{-1}Y}{3} \right) \right),
\end{eqnarray}
where
\begin{equation}
X= 4+2\sqrt{3}\beta_m\,\,\,\,\,  \text{and}\,\,\,\,\, Y=-\frac{3 \sqrt{3} \beta _m+5}{\sqrt{2} \left(\sqrt{3} \beta _m+2\right){}^{3/2}}.
\end{equation}

\begin{figure}[h!]
\begin{center}
\includegraphics[scale=0.6]{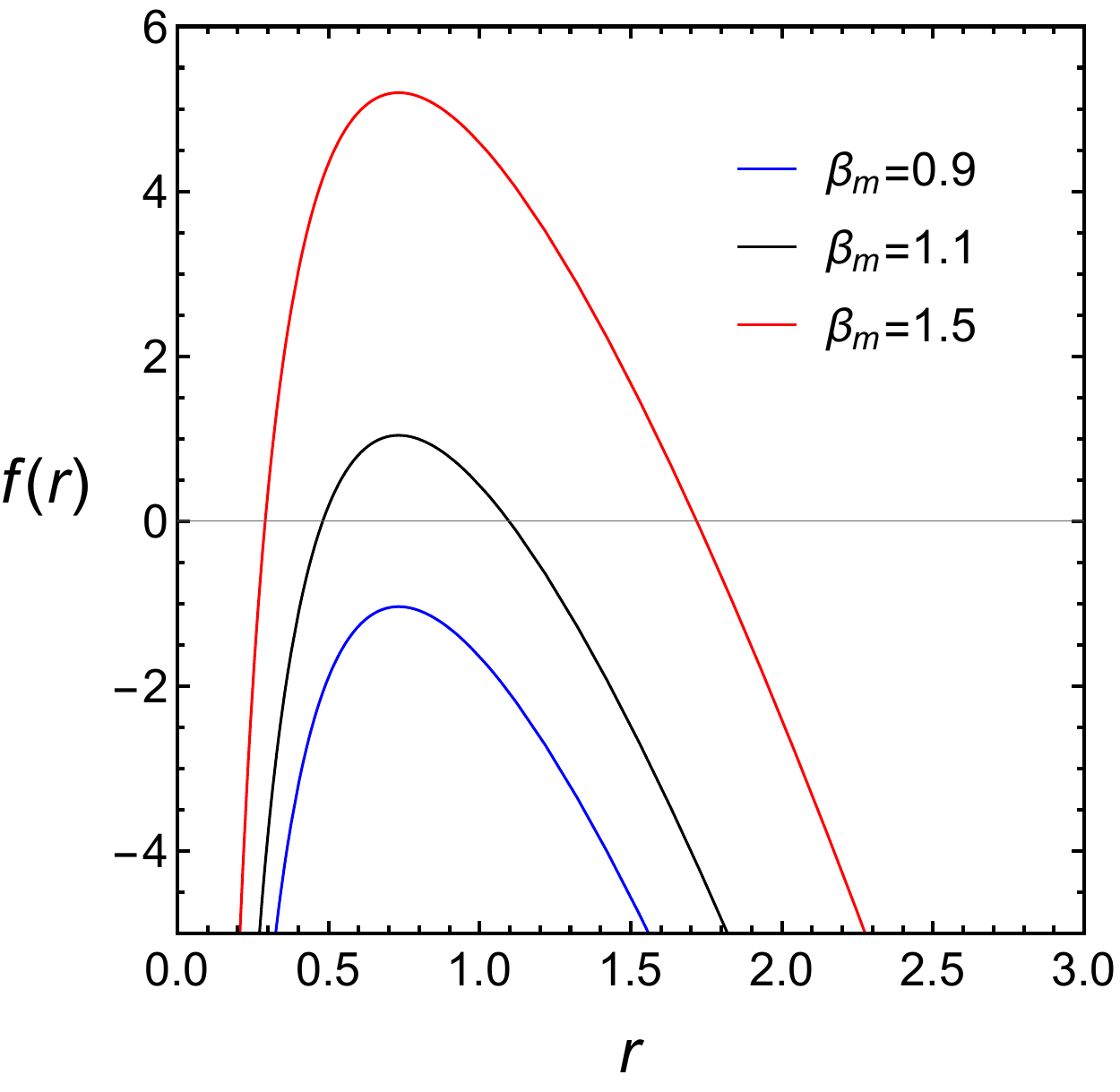}
\end{center}
{\caption{ Plot of $f(r)$ using different values of $\beta_m$, with $M=1$ and $c_2 = -1$}\label{fMGdS}}
\end{figure}
One can see that we now have two parameters, $c_2$ and $\beta_m$ to control the behaviour of the horizons. The parameter $c_2$ controls the strength of the graviton mass or the cosmological constant, while $\beta_m$ controls the existence of the horizons. For $0 < \beta_m <1$, there are no horizons, while for $\beta_m > 1$, there are two horizons, and then such two horizons become closer and closer when $\beta_m$ approaches $1$, and thus such two horizons merge together at $\beta_m = 1$ as shown in Fig. \ref{fMGdS}.

It is useful to emphasize here that our choice, $c_1 = 6 (M c_2^2)^{1/3}$, provides only a class of conditions to characterize the existence of the horizons. It is not valid in general. For example, for $c_0 =0$ corresponding to $\beta_m=0$, it is still possible to find the parameter space for $c_2$ and $c_1$ to have two horizons. Even though this choice and the this set of parameters ($c_2, \beta_m$), provide a loss of generality of parameter space, it provides us with a qualitative way to analyze the effects of the horizon structure on the potential and the greybody factor. This will be explicitly shown later in Sec. \ref{eom} and Sec. \ref{bound}.

It is important to note that the existence of parameters, $c_1$ and $c_0$, is characterized by the structure of the dRGT massive gravity theory, which provides an additional part to the usual dS black string solution \cite{Lemos:1994xp,Lemos:1994fn}. From Eq. (\ref{f}), one can see that without these parameters ($c_2 <0, c_1=0, c_0=0$), it is not possible to have a horizon since $f$ is always negative; therefore, it is not possible to investigate the thermodynamics of the black string or find the greybody factor for the dS black string solution. This is a crucial issue for the dRGT massive gravity black string solution, and we will investigate this issue in the next section.

\subsection{Asymptotic AdS solution}
For the asymptotic AdS solution, $c_2 >0$, one can use the same strategy as in the previous subsection, in finding two extremas when $f' = 0$. As a result, these two extremas can be written as
\begin{eqnarray}
r_{ex1} &=& \left(\frac{M}{c_2}\right)^{1/3},\\
r_{ex2} &=& (1+\sqrt{3})\left(\frac{M}{c_2}\right)^{1/3}.
\end{eqnarray}
Following the same step, the function $f$ at the extrema can be written as
\begin{eqnarray}
f(r_{ex1}) &=& c_0 - 6 \sqrt{3} c_2 \left(\frac{M}{c_2}\right)^{2/3},\\
f(r_{ex2}) &=& c_0 - 9 c_2 \left(\frac{M}{c_2}\right)^{2/3}.
\end{eqnarray}
In order to see the structure of the horizons, let us define a parameter to parametrize the existence of three horizon as
\begin{eqnarray}
c_0 &=& \beta_m 6 \sqrt{3} c_2 \left(\frac{M}{c_2}\right)^{2/3},
\end{eqnarray}
where the condition for having three horizon is
\begin{eqnarray}
\frac{\sqrt{3}}{2}<\beta_m <1.
\end{eqnarray}
By substituting these parameters into $f$, and then finding the solution of $f =0$ for $r$, one obtains three horizons as follows
\begin{eqnarray}
r_1 &=& \frac{2 (c_2^2 M)^{1/3}}{c_2}\left( 1- \sqrt{x} \sin\left(\frac{\cos^{-1}y}{3} + \frac{\pi}{6}\right) \right),\\
r_2 &=& \frac{2 (c_2^2 M)^{1/3}}{c_2}\left( 1- \sqrt{x} \cos\left(\frac{\cos^{-1}y}{3} + \frac{\pi}{3} \right) \right),\\
r_3 &=& \frac{2 (c_2^2 M)^{1/3}}{c_2}\left( 1+ \sqrt{x} \cos\left(\frac{\cos^{-1}y}{3} \right) \right),
\end{eqnarray}
where
\begin{equation}
x= 4-2\sqrt{3}\beta_m\,\,\,\,\,  \text{and}\,\,\,\,\, y=\frac{5-3 \sqrt{3} \beta _m}{\sqrt{2} \left(2-\sqrt{3} \beta _m\right){}^{3/2}}.
\end{equation}

\begin{figure}[h!]
\begin{center}
\includegraphics[scale=0.5]{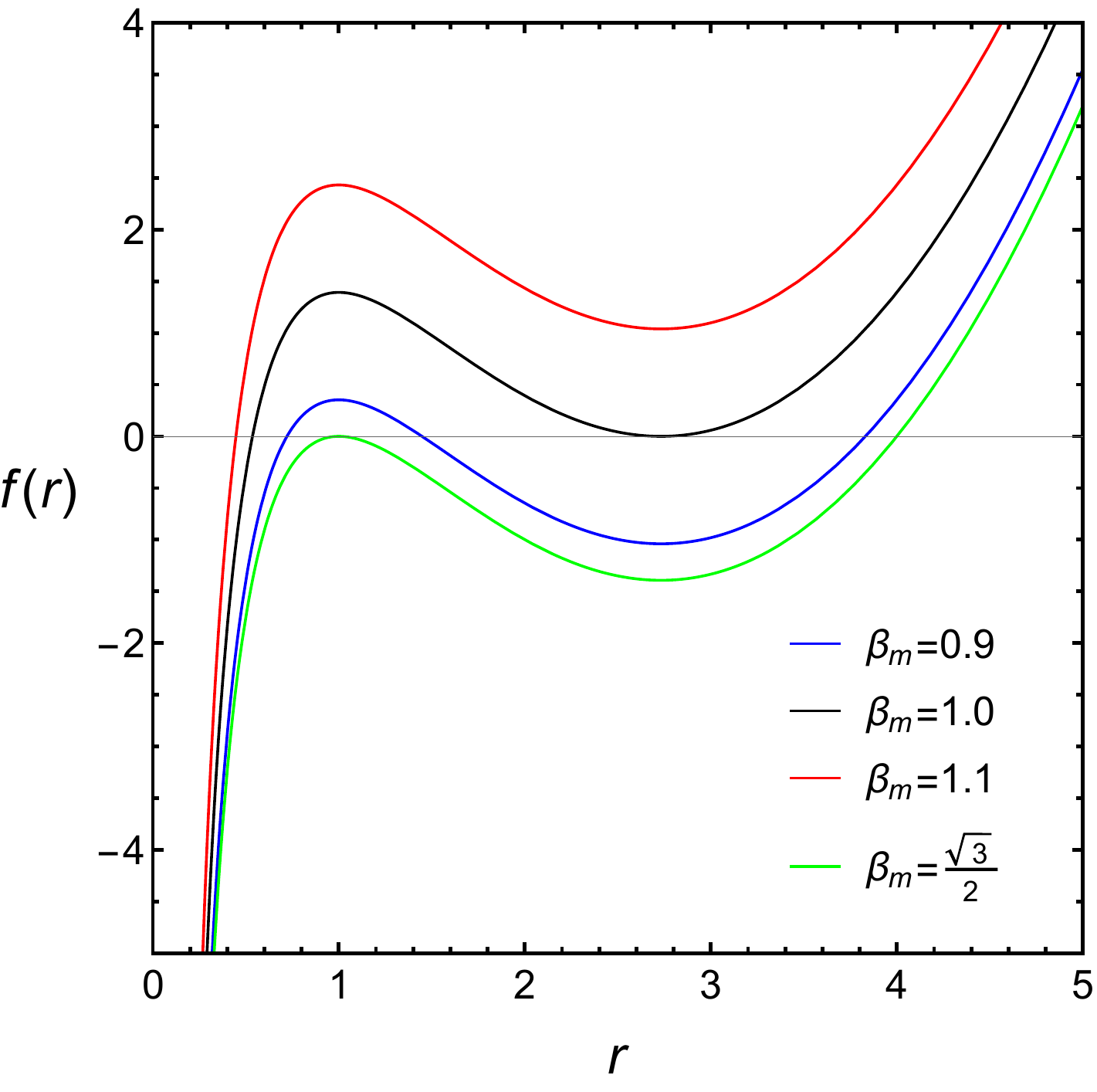}
\end{center}
{\caption{ Plot of $f(r)$ using different values of $\beta_m$, with $M=1$ and $c_2 = 1$}\label{fMGAdS}}
\end{figure}

As we have analyzed in the previous subsection, we recover the usual AdS black string solution by setting $c_2 > 0$ and $c_0 = c_1 =0$. In this case, it is found that there exist only one horizon. Therefore, the crucial difference is characterized by the existence of $c_1$ and $c_0$, which are now re-parametrized by only one parameter $\beta_m$.  As we have seen in Fig. \ref{fMGAdS}, one can obtain three horizons for $\sqrt{3}/2 < \beta_m < 1$. For $\beta_m = \sqrt{3}/2$, the first and the second horizons are merged together, while when $\beta_m = 1$, the second and the third horizons are merged together, with two horizons for these two specific cases. Finally, one horizon can exist for $0 < \beta_m < \sqrt{3}/2$ (third horizon) and $\beta_m > 1$ (first horizon). This behaviour can be seen explicitly in Fig. \ref{fMGAdS}.

Note that even though we leave only two parameters for characterizing the behaviour of the horizon structure, this is very useful for the analytical investigation of how the horizon structure influences the potential form and also the greybody factor. We will show this analysis in the next two sections.

\section{Equations of motion of the massless scalar field}\label{eom}
\begin{figure}[h!]
\begin{center}
\includegraphics[scale=0.52]{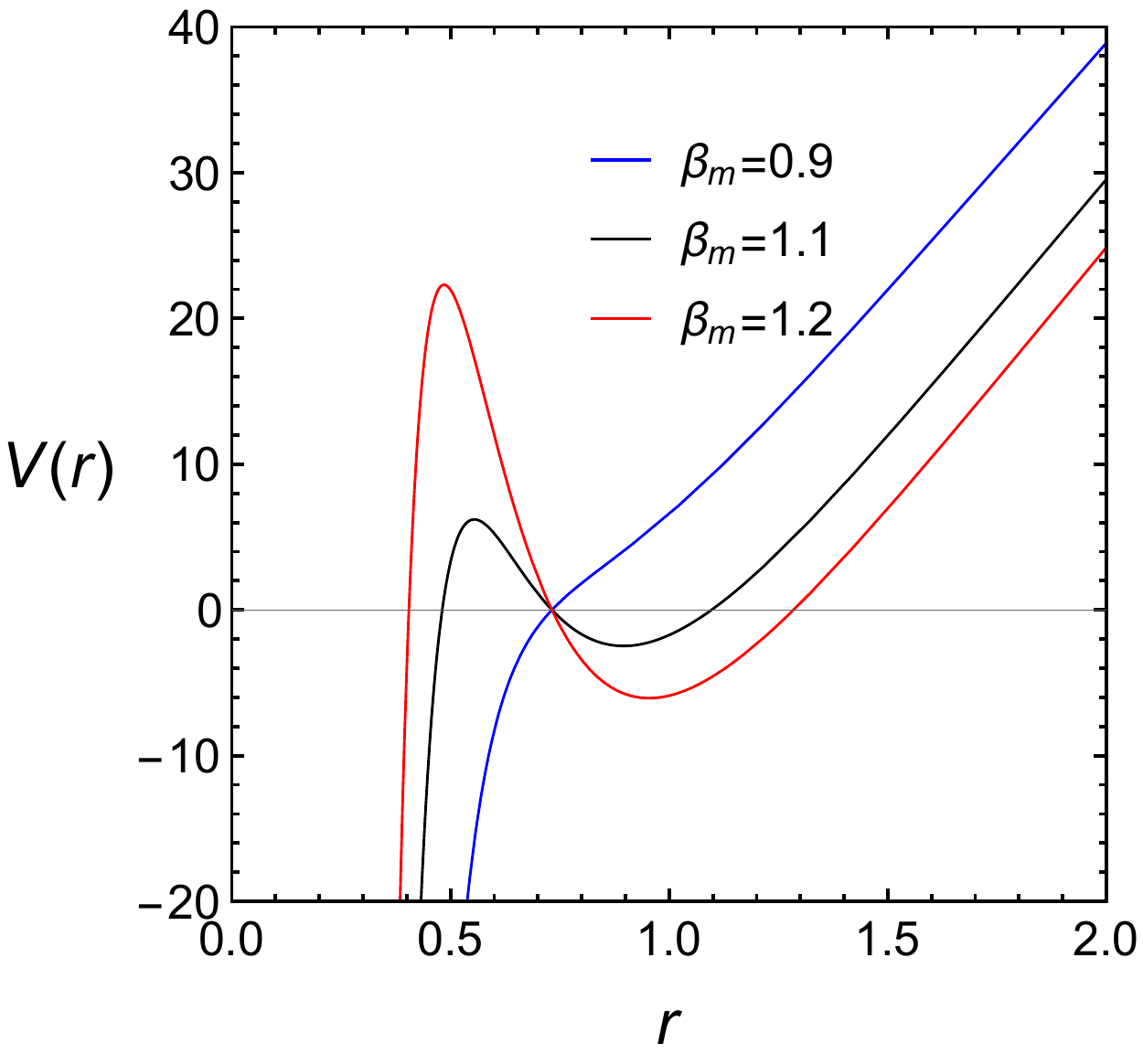}\qquad
\includegraphics[scale=0.5]{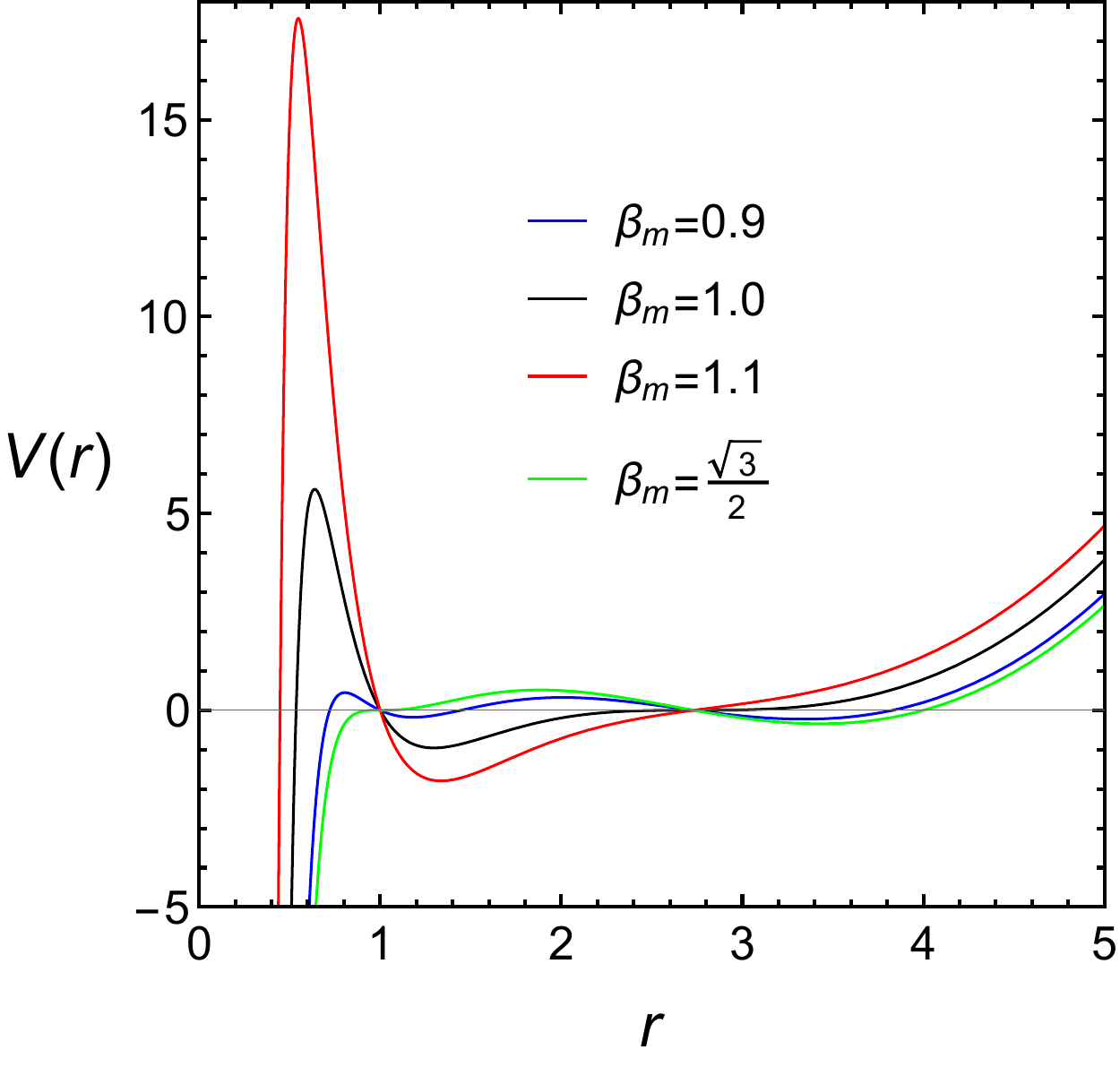}
\end{center}
{\caption{The left panel shows the potential for the asymptotic dS solution with $\ell = 0$, $c_2 = -1$,  $M = 1$. The right panel shows the potential for the asymptotic AdS solution with $\ell = 0$, $c_2 = 1$,  $M = 1$.}\label{pott}}
\end{figure}

In this work, we are interested in a massless uncharged scalar field emitted from the dRGT black string as Hawking radiation. The equation of motion, which describes the motion of the massless uncharged scalar field, is the Klein-Gordon equation,

\begin{equation}
\frac{1}{\sqrt{-g}}\partial_{\mu}\left(\sqrt{-g}g^{\mu\nu}\partial_{\nu}\Phi\right) = 0.
\end{equation}
By using the solution of the physical metric in Eq. (\ref{solutiong}), the solutions  can be separated in the form
\begin{equation}
\Phi(t, r, \Omega) = T(t) Y(\varphi,z)\frac{\psi(r)}{r},
\end{equation}
where $T = e^{\pm i \omega t }$ is the oscillating function and $Y$ satisfies the equation,
\begin{equation}
\frac{\partial^2 Y}{\partial\varphi^2}+ \frac{1}{\alpha_g^2}\frac{\partial^2 Y}{\partial z^2} = - \ell(\ell+1)Y.
\end{equation}
The radial part of the Klein-Gordon equation is
\begin{equation}
\frac{d^{2}\psi(r)}{dr_{*}^{2}} + \left[\omega^{2} - V(r)\right]\psi(r) = 0,
\end{equation}
where $r_{*}$ is the tortoise coordinate defined by
\begin{equation}
\frac{dr_{*}}{dr} = \frac{1}{f(r)}\label{tortoise}
\end{equation}
and $V(r)$ is the potential given by
\begin{equation}
V(r) = f(r)\left[\frac{\ell(\ell + 1)}{r^{2}} + \frac{f'(r)}{r}\right].\label{poten}
\end{equation}
Surprisingly, the equation of motion for the radial part is in the same form, even though we used the cylindrical coordinates instead of the spherical coordinates. This allows us to perform the investigation for the greybody factor in the same fashion as usually done in spherical coordinates. Moreover, since the form of the radial equation is still in the form of Schrodinger-like equation, one can perform the analysis of the effect of the potential form on the transmission amplitude similar to one in quantum mechanics. It is important to note that the leading contribution to the transmission amplitude or the greybody factor is the mode $\ell = 0$, since the larger the value of $\ell$, the higher the value of the potential and the more difficult it is for the wave to transmit. This behaviour is also common in the spherical symmetry case.  As a result, we will restrict our attention to the case $\ell = 0$, and then the potential becomes $V = f'f/r$. In order to see the behaviour of the potential in terms of the massive graviton parameters, one can substitute $f$ from equation (\ref{f}), and then reparametrize the parameters in terms of $\beta_m$ and $c_2$. As a result, by fixing $c_2$, and then varying $\beta_m$, the behaviour of the potential in both the asymptotic dS and the asymptotic AdS solutions can be illustrated in   Fig. \ref{pott}. From the left panel of this figure (the asymptotic dS case), one can see that the potential becomes lower when the parameter $\beta_m$ approaches $1$. In other words, when the horizons become closer, the potential becomes lower and lower. This gives a hint to us that the greybody factor bound will be higher when the horizons become closer. This analysis is also valid for the asymptotic AdS case. We will consider this analysis in detail in the next section.
%\begin{equation}
%V(r) = f(r)\left[\frac{\ell(\ell + 1)}{r^{2}} + 2 c_2 + \frac{4M}{\alpha_{g}r^{3}} - \frac{c_{1}}{r}\right].\label{poten2}
%\end{equation}

\section{The rigorous bounds on the greybody factors}\label{bound}
There are many methods to calculate the greybody factor such as the matching technique and the WKB approximation \cite{Parikh,Fleming,Fernando,Lange,Kim,Escobedo,Harmark,Kanti,Dong}. In this present work, we will focus on the method that does not use such approximation, namely, the rigorous bound on the greybody factor. The advantage of this method is that it provides us with a better way to analyze the greybody factor qualitatively. Then, the influence of the potential form on the greybody factor can be explored. The rigorous bounds on the greybody factors are given by
\begin{equation}
T \geq \text{sech}^{2}\left(\int_{-\infty}^{\infty}\vartheta dr_{*}\right),
\end{equation}
where
\begin{equation}
\vartheta = \frac{\sqrt{[h'(r_{*})]^{2} + \left[\omega^{2} - V(r_{*}) - h^{2}(r_{*})\right]^{2}}}{2h(r_{*})}
\end{equation}
and $h(r_{*})$ is a positive function satisfying $h(-\infty) = h(\infty) = \omega$. See \cite{1D} for more details. We select $h = \omega$. Therefore,
\begin{equation}
T \geq \text{sech}^{2}\left(\frac{1}{2\omega}\int_{-\infty}^{\infty}|V|dr_{*}\right).\label{T}
\end{equation}
From equation (\ref{poten}), together with $f$ in Eq. (\ref{f}), the potential is
\begin{equation}
V(r) = f(r)\left[\frac{\ell(\ell + 1)}{r^{2}} + 2 c_2 + \frac{4M}{r^{3}} - \frac{c_{1}}{r}\right],
\end{equation}
where $f(r)$ is given by equation (\ref{f}). From equation (\ref{tortoise}), the rigorous bound on the greybody factor given by equation (\ref{T}) becomes
\begin{equation}
T \geq T_b = \text{sech}^{2}\left(\frac{1}{2\omega}\int_{r_{H}}^{R_{H}}\frac{|V|}{f(r)}dr\right) = \text{sech}^{2}\left(\frac{A_{\ell}}{2\omega}\right),\label{Tb}
\end{equation}
where
\begin{equation}
A_{\ell} = \int_{r_{H}}^{R_{H}}\frac{|V|}{f(r)}dr =\int_{r_{H}}^{R_{H}}\left|\frac{\ell(\ell + 1)}{r^{2}}+\frac{f'}{r}\right|dr.
\end{equation}
As we know, the function $\text{sech}^{2}$ is maximum at $\text{sech}^{2}(0)$, so that the function $A_{\ell}$ must be close to zero in order to obtain the higher value of the bound $T_b$.
Therefore, one can ignore the contribution from $A_{\ell}$ with $\ell \geq 1$. Now, let us consider $A_{0}$, which can be written as
\begin{equation}
A_{0}  =\int_{r_{H}}^{R_{H}}\left|\frac{f'}{r}\right|dr.
\end{equation}

\begin{figure}[h!]
\begin{center}
\includegraphics[scale=0.55]{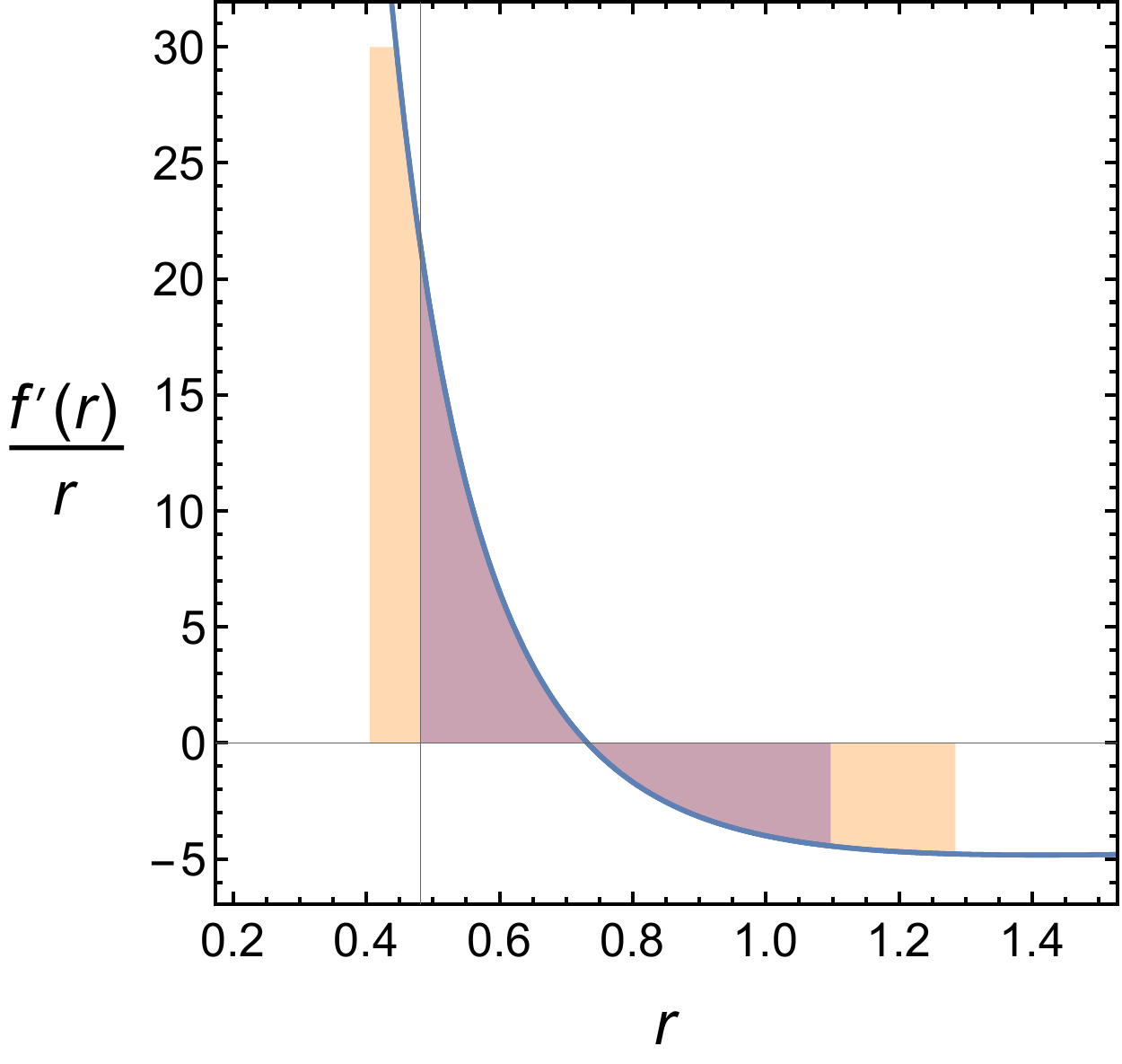}
\end{center}
{\caption{Plot of $f'/r$ using different values of $\beta_m$, with $M=1$ and $c_2 = 1$. The blue filled area corresponds to $\beta_m = 1.1$ and the orange filled area corresponds to $\beta_m = 1.2$}\label{A0}}
\end{figure}

\begin{figure}[h!]
\begin{center}
\includegraphics[scale=0.55]{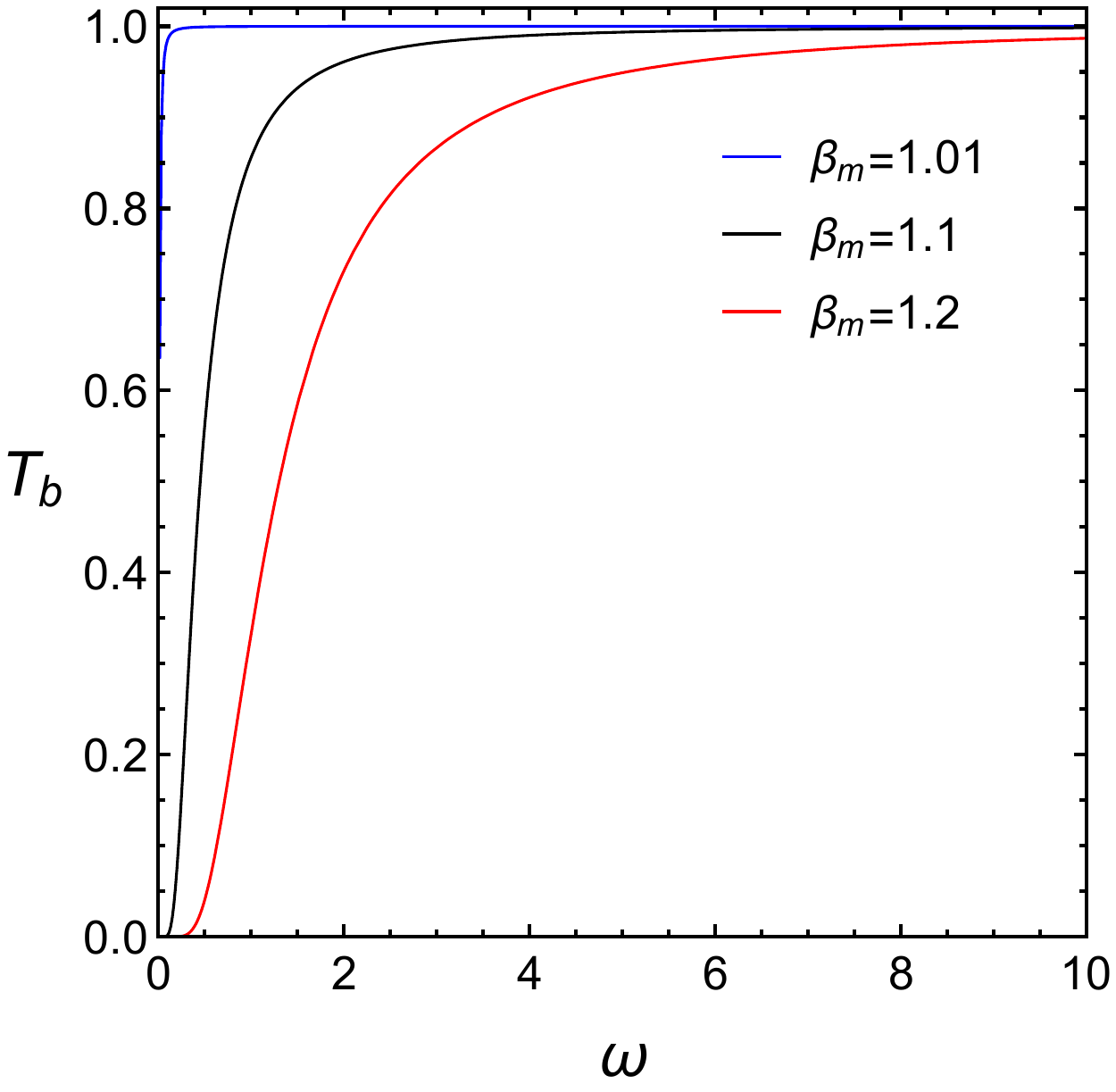}
\end{center}
{\caption{ Plot of $T_b$ using different values of $\beta_m$, with $\ell = 0,M=1$ and $c_2 = -1$.}\label{TmgdS}}
\end{figure}

One can see that $A_0$ is the area filled by function $f'/r$. Since the function $f'/r$ does not depend on $c_0$, it therefore does not depend on $\beta_m$. After fixing $c_1$ and $c_2$, $f'/r$ is still the same function. Therefore, the filled area is different by the limit of integration as shown in Fig \ref{A0}. This can also be seen from Fig. \ref{fMGdS} as the value of $\beta_m$ is close to $1$ where two horizons are sunk together. This analysis is also confirmed by using a numerical method as shown in Fig. \ref{TmgdS}. Moreover, this behaviour is also consistent with the shape of the potential as illustrated in the left panel of Fig. \ref{pott}. From this figure, it can be inferred that if the potential is higher,  the value of the transmission amplitude is lower.

\begin{figure}[h!]
\begin{center}
\includegraphics[scale=0.44]{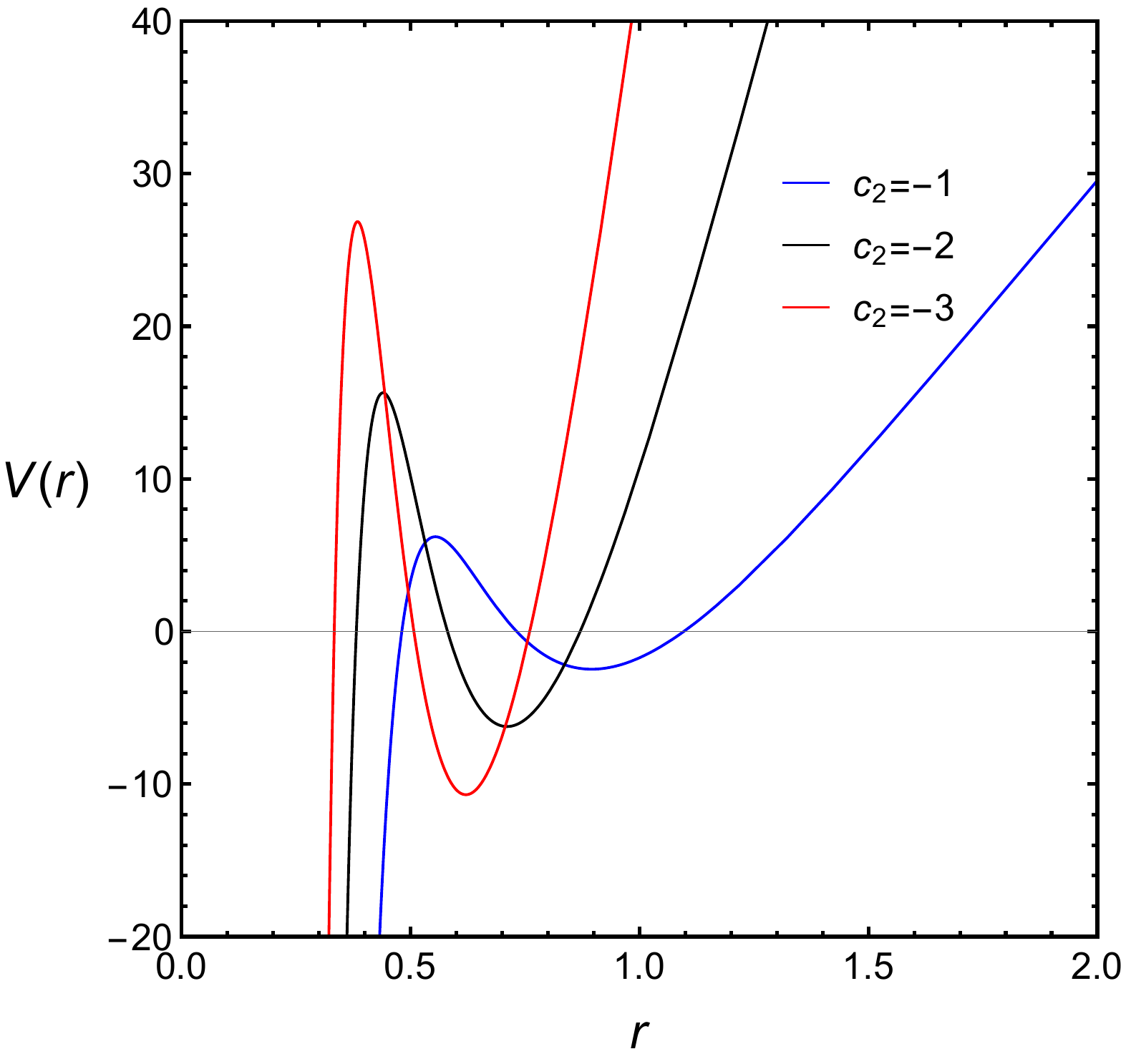}\qquad
\includegraphics[scale=0.48]{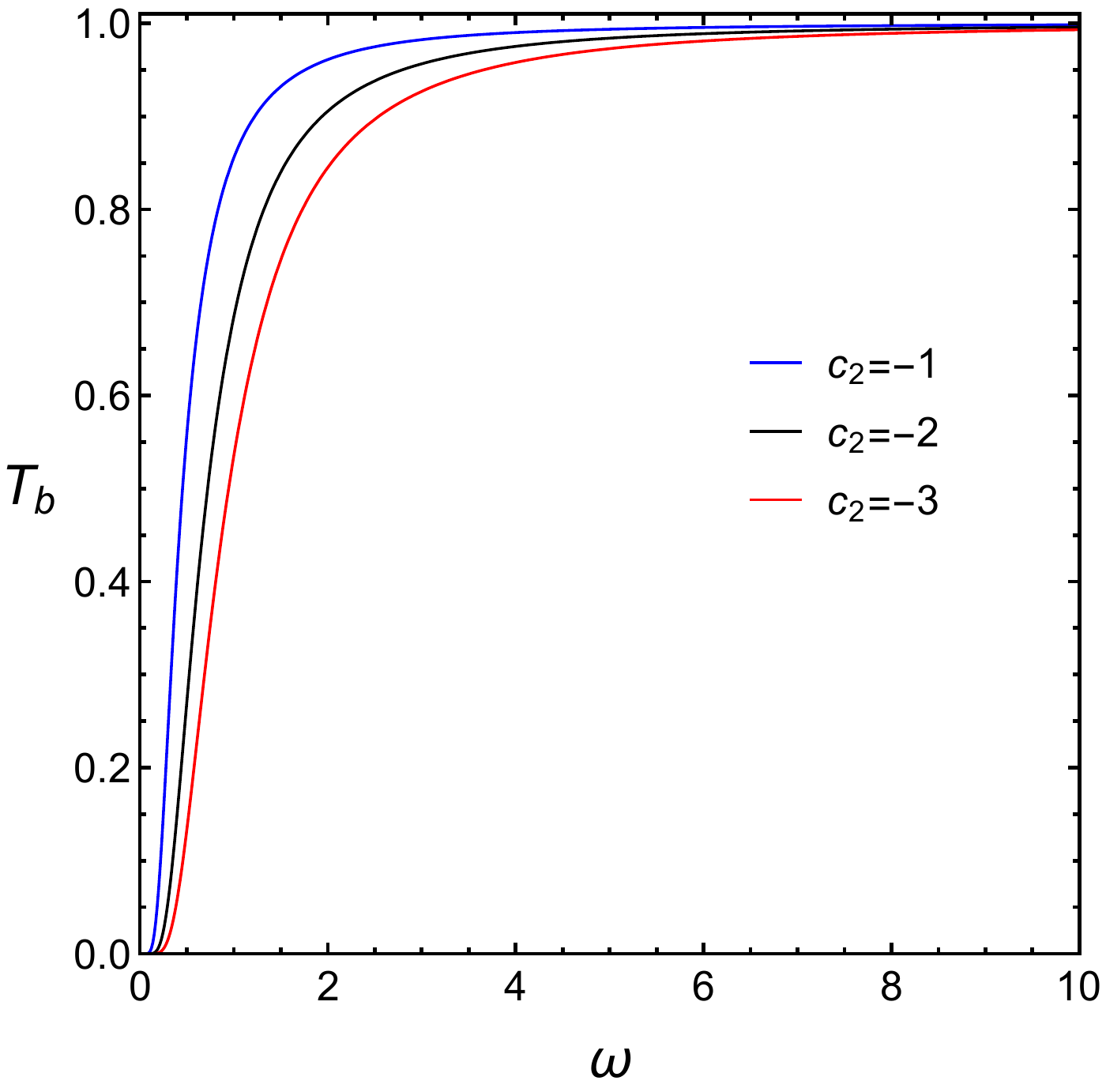}
\end{center}
{\caption{These plots show the horizon structure, shape of the potential and the greybody factor bound for different values of $c_2$, with $M = 1$ and $\beta_m=1.1$.}\label{consitencyc2dS}}
\end{figure}

In order to find the effect of parameter $c_2$, one can fix the parameter as $\beta_m = 1.1$. As a result, the shape of the potential will control the greybody factor bound. This is similar to one in quantum theory, where the higher the potential, the lower the transmission amplitude and then the lower the greybody factor bound. This consistency is shown in Fig. \ref{consitencyc2dS}. From these figures, one can see that the larger the value of $|c_2|$, the higher the value of the potential and then the lower the value of the greybody factor bound.

Now, let us consider the asymptotic AdS solution. As we have discussed, it is possible to obtain the 3 horizons for this kind of solutions. In this case, one may have to suppose the place of the observer. As a result, we can divide  our consideration into two cases; the observer being between the first and the second horizons, and the other being between the second and the third horizons. From Fig \ref{fMGAdS}, one finds that three horizons exist if $\sqrt{3}/2 < \beta_m < 1$. For $\beta_m = \sqrt{3}/2$, the first and the second horizons are sunk together, and for $\beta_m = 1$, the second and the third horizons are sunk together.

\begin{figure}[h!]
\begin{center}
\includegraphics[scale=0.5]{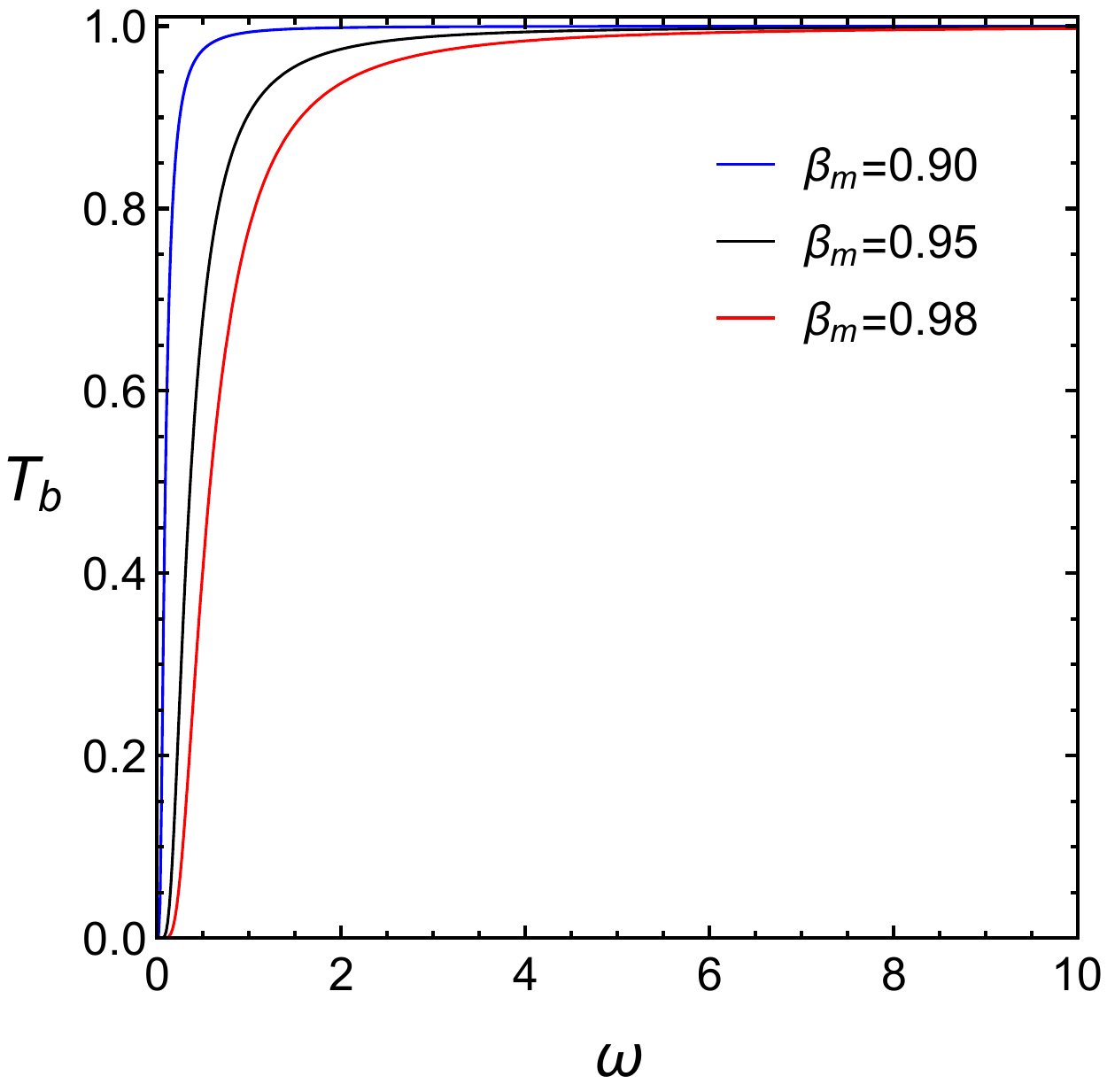}\qquad
\includegraphics[scale=0.5]{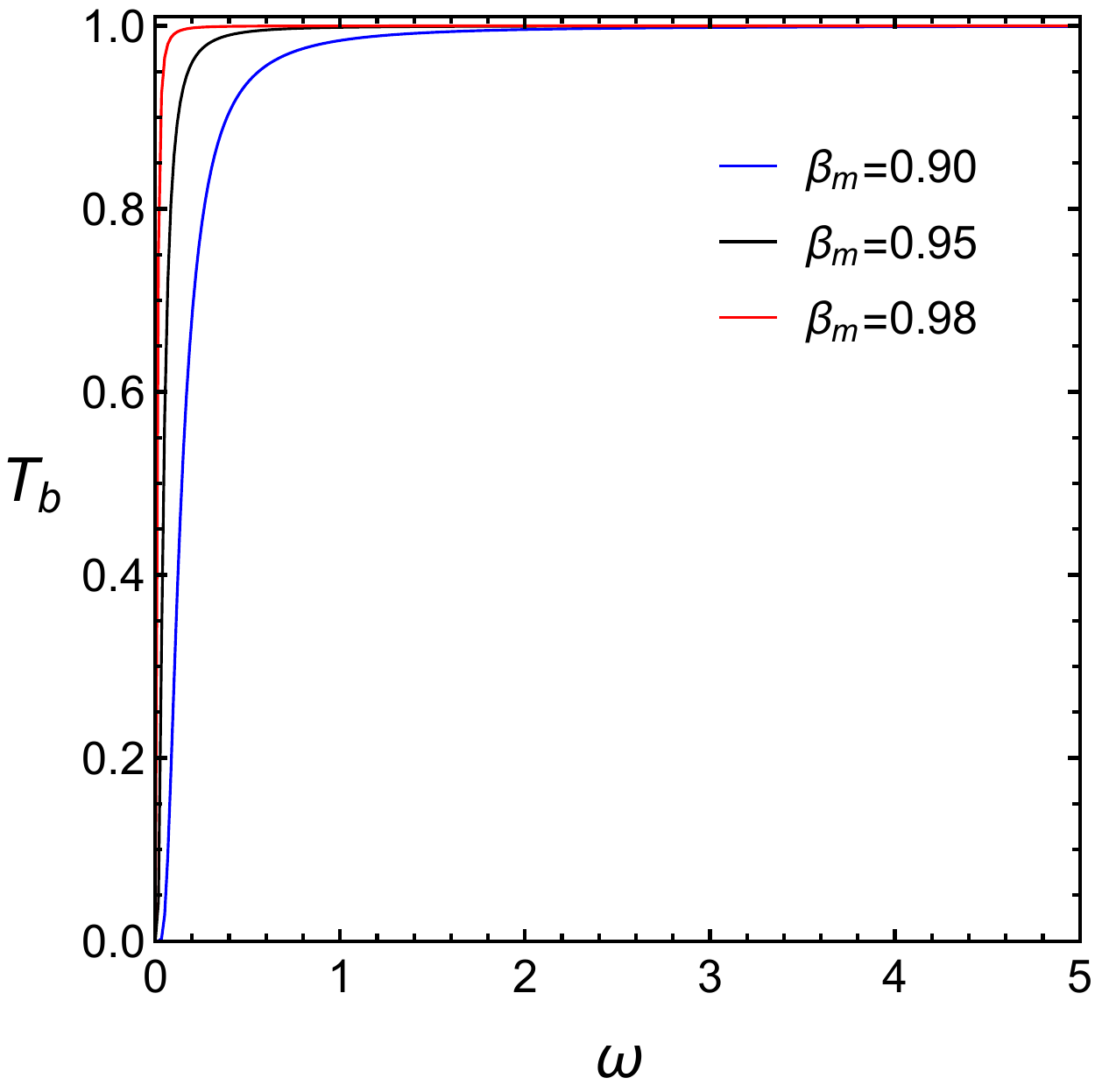}
\end{center}
{\caption{Plot of $T_b$ using different values of $\beta_m$, with $\ell = 0,M=1$ and $c_2 = 1$. The left panel is for the one between the first and the second horizon, and the right panel is for one between the second and the third horizon}\label{TmgAdS}}
\end{figure}

By fixing $c_2$, one can still use the same analysis as done in the asymptotic dS case, where the greybody factor bound depends crucially on the distance between the horizons. These can be seen explicitly in Fig \ref{TmgAdS}.

\begin{figure}[h!]
\begin{center}
\includegraphics[scale=0.4]{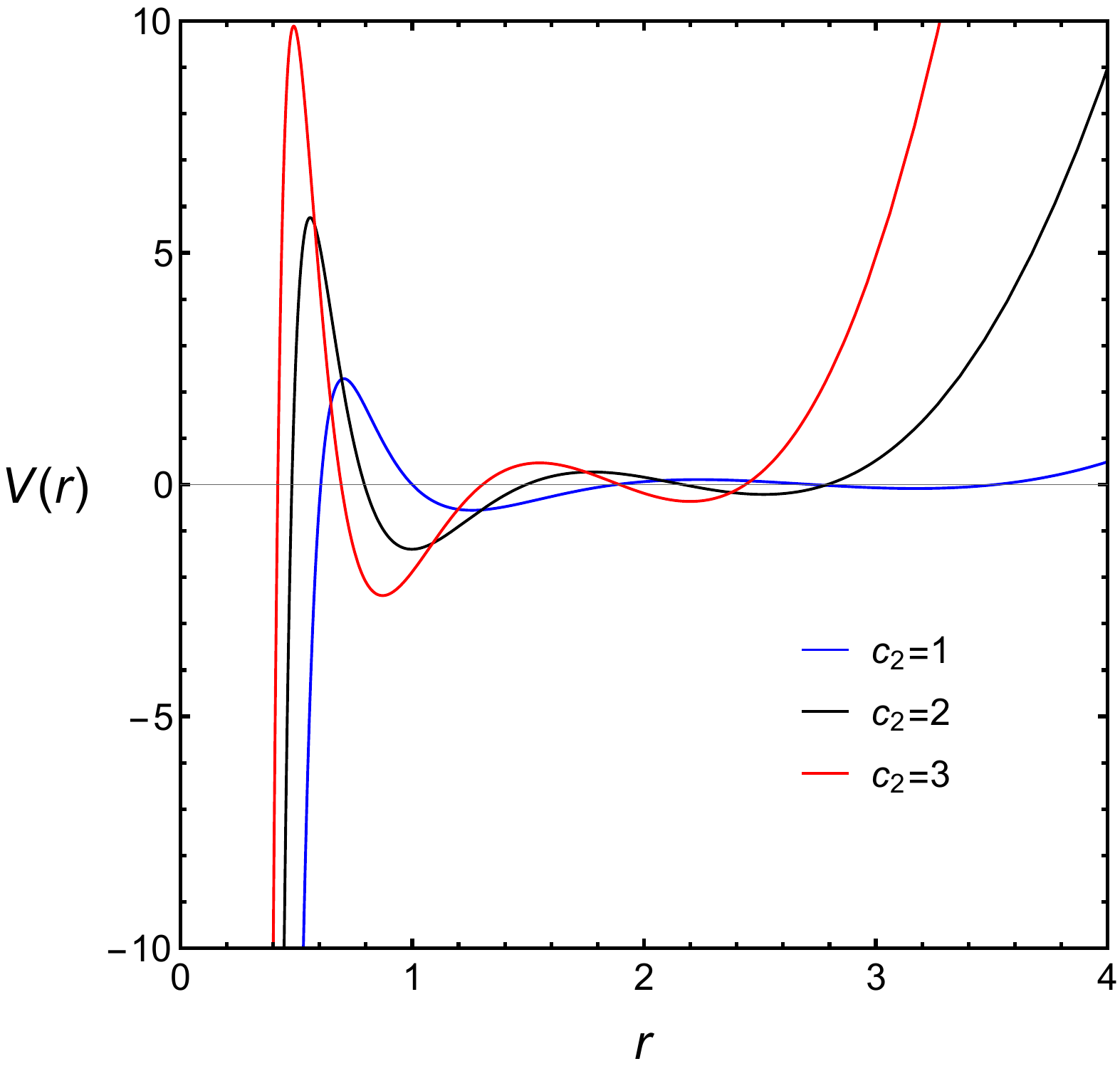}\qquad
\includegraphics[scale=0.45]{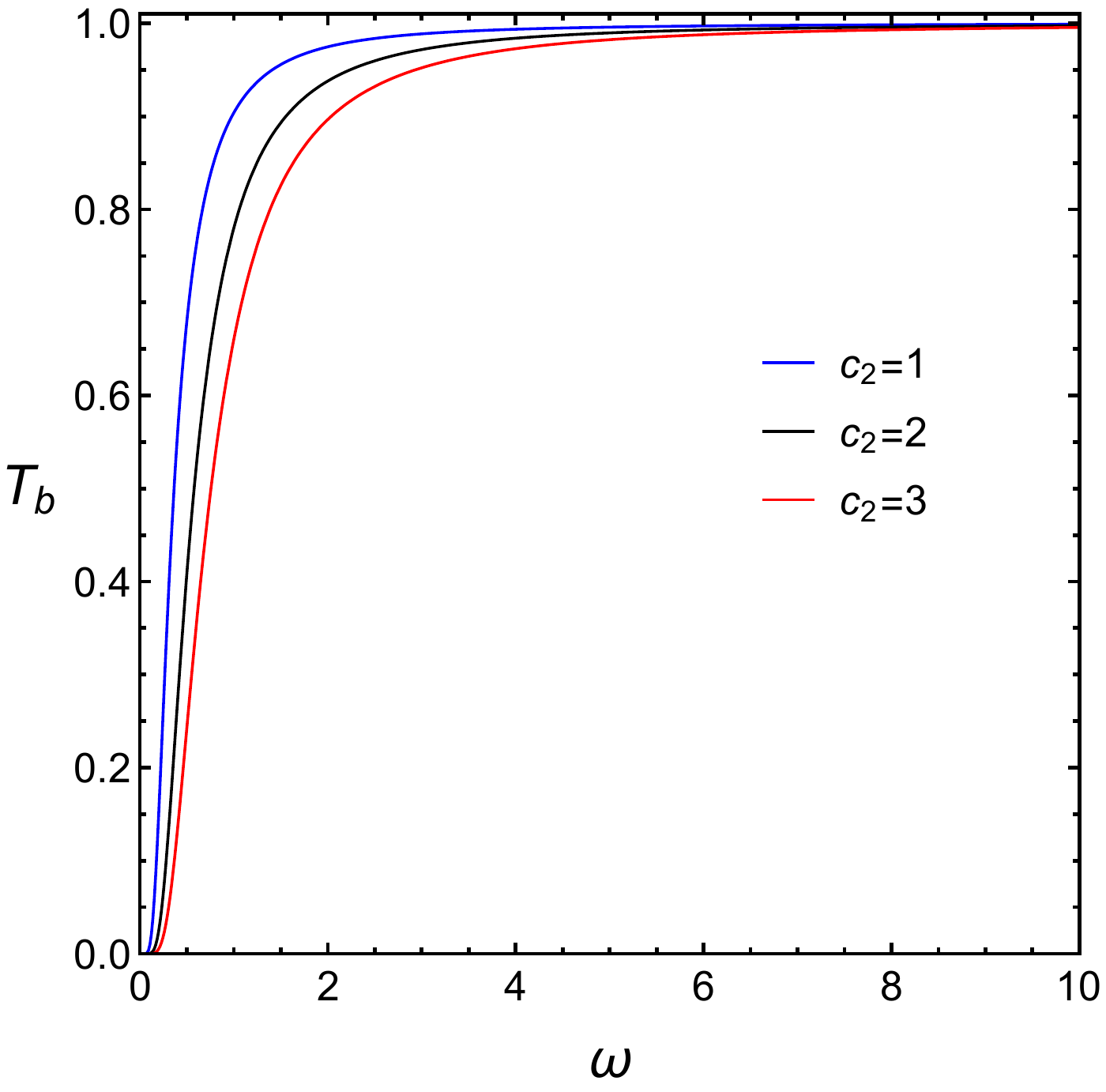}
\end{center}
{\caption{These plots show the horizon structure, shape of the potential and the greybody factor bound for different values of $c_2$, with $M = 1$ and $\beta_m=0.95$.}\label{consitencyc2AdS}}
\end{figure}

Now, let us fix the parameter $\beta_m$. As we have analyzed above, the greybody factor bound crucially depends on the maximum value of the potential; the higher the value of the potential, the more difficult it is for the waves to be transmitted and then the lower the bound of the greybody factor. This behavior is shown explicitly in Fig. \ref{consitencyc2AdS}, where the shape of the potential is in the left panel and the corresponding greybody factor bound is in the right panel.

\section{Conclusion}\label{con}
In this paper, we investigated the greybody factor of the black string in dRGT massive gravity theory by using the rigorous bound. In order to properly study the dRGT black string, we first investigated the horizon structures of the dRGT black string. We defined the new model parameter $\beta_m$ to characterize the existence of the horizons. The results show that, for the asymptotic dS solution, there are two horizons when $\beta_{m} > 1$, and for the asymptotic AdS solution, there are three horizons when $\sqrt{3}/2 < \beta_{m} < 1$. By considering a massless uncharged scalar field emitted from the dRGT black string as Hawking radiation, the Schrodinger-like equation is obtained for the radial part of the solution. As a result, this allows us to consider the behaviour of the potential for investigating the greybody factor. It is found that the height of the potential becomes lower when the parameter $\beta_m$ approaches $1$ for the asymptotic dS solution, while $\beta_m$ approaches $1, \sqrt{3}/2$ for the asymptotic AdS solution where two horizons are merged together. Moreover, the rigorous bounds on the greybody factors have also been calculated. It is found that the greybody factor bound can be qualitatively analyzed by using the following form of potential; the higher the value of the potential, the more difficult it is for the waves to be transmitted and then the lower the bound of the greybody factor. This result is valid for both the asymptotic AdS solution and the asymptotic dS solution, and also checked by numerical method. Since our analysis/results are similar to ones in quantum mechanics, it provides us with an easier way to deal with the quantum nature of black holes or black strings, even though a complicated form of spacetime is considered.

\section*{Acknowledgement}
This project was funded by the Ratchadapisek Sompoch Endowment Fund, Chulalongkorn University (Sci-Super 2014-032), by a grant for the professional development of new academic staff from the Ratchada pisek Somphot Fund at Chulalongkorn University, by the Thailand Research Fund (TRF), and by the Office of the Higher Education Commission (OHEC), Faculty of Science, Chulalongkorn University  (RSA5980038). PB was additionally supported by a scholarship from the Royal Government of Thailand. TN was also additionally supported by a scholarship from the Development and Promotion of Science and Technology Talents Project (DPST). PW was supported by the Thailand Research Fund (TRF) through grant no. MRG6180003 and partially supported by the ICTP through grant no. OEA-NT-01.

\end{document}